\documentclass[twocolumn]{llncs}

\usepackage[left=2cm, right=2cm, top=3cm, bottom=3cm]{geometry}

\usepackage{amsmath}
\usepackage{amssymb}
\usepackage{mathtools}
\usepackage{bm}
\usepackage{cite}
\usepackage{todonotes}
\usepackage{acro}
\usepackage{tikz}
\usetikzlibrary{angles,decorations.pathreplacing,decorations.text,calligraphy}
\usepackage{braket}

\usepackage{listings}
\usepackage{subcaption}
\usepackage{cleveref}

\usepackage{tabularx}
\usepackage{booktabs}

\DeclareAcronym{cfg}{
  short=CFG,
  long=control flow graph,
}
\DeclareAcronym{ddg}{
  short=DDG,
  long=data-dependence graph,
}
\DeclareAcronym{ipe}{
  short=IPE,
  long=Iterative phase estimation,
}
\DeclareAcronym{ir}{
  short=IR,
  long=intermediate representation,
}
\DeclareAcronym{hqcc}{
  short=HQCC,
  long=heterogeneous quantum-classical computation,
}
\DeclareAcronym{nisq}{
  short=NISQ,
  long=noisy intermediate-scale quantum,
}
\DeclareAcronym{qct}{
  short=QCT,
  long=quantum calculation time,
}
\DeclareAcronym{qin}{
  short=QIN,
  long=quantum instruction number,
}
\DeclareAcronym{qvm}{
  short=QVM,
  long=quantum virtual machine,
}
\DeclareAcronym{qpu}{
  short=QPU,
  long=quantum processing unit,
}

\title{Optimization of Hybrid Quantum-Classical Algorithms\thanks{This work is based on the master thesis of the first author~\cite{Remme2025}.}
}

\author{Lian Remme\inst{1} \and
Alexander Weinert\inst{1} \and
Andre Waschk\inst{1}}
\authorrunning{Remme, Weinert, and Waschk}
\institute{German Aerospace Center (DLR), Cologne, Germany
\email{\{lian.remme,alexander.weinert,andre.waschk\}@dlr.de}}

\begin{document}

\maketitle

\begin{abstract}

Quantum computers do not run in isolation; rather, they are embedded in quantum-classical hybrid architectures.
In these setups, a quantum processing unit communicates with a classical device in near-real time.
To enable efficient hybrid computations, it is mandatory to optimize quantum-classical hybrid code. 
To the best of our knowledge, no previous work on the optimization of hybrid code nor on metrics for which to optimize such code exists.

In this work, we take a step towards optimization of hybrid programs by introducing seven optimization routines and three metrics to evaluate the effectiveness of the optimization.
We implement these routines for the hybrid quantum language Quil and show that our optimizations improve programs according to our metrics.
This lays the foundation for new kinds of hybrid optimizers that enable real-time collaboration between quantum and classical devices.

\end{abstract}


\section{Introduction}
\label{sec:introduction}

In recent years, quantum computing has advanced from a theoretical possibility researched in foundational physics to the development of working quantum computers on which quantum algorithms can be executed.
Mainly, quantum computers serve as stand-alone executors of \emph{pure (quantum) programs}.
For this, one first encodes the program as a quantum circuit~\cite{qiskit_quantum_circuit,pennylane_quantum_circuit,cirq_circuits}.
A CPU then sends this circuit to a \ac{qpu}, which executes it and sends the results of the execution back to the CPU~\cite{knill1996conventions}.
This mode of execution only requires non-real-time communication capabilities between the CPU and the QPU.

Some algorithms, in contrast, require \emph{real-time communication}~\cite{bennett1993teleporting,lubinski2022advancing,paetznick2013repeat,fu2023proto,dobvsivcek2007arbitrary,fu2021quingo,rossi2022using} between the CPU and the QPU, i.e.\,communication within the coherence time of the qubits.
In this execution model, the circuit to be executed is not fixed in advance.
Instead, the CPU obtains intermediate computation results during the execution of the circuit and may influence subsequent execution steps.
This allows to, e.g., reduce the quantum resources required by an algorithm~\cite{beauregard2002circuit,rossi2022using,dobvsivcek2007arbitrary,fu2021quingo} or to check whether a desired quantum state has been created~\cite{lubinski2022advancing,knill2004fault,bravyi2005universal,fowler2012surface,paetznick2013repeat,fu2023proto}.
We call quantum programs that require real-time communication between the CPU and the QPU \emph{real-time hybrid (quantum) programs}.

Modern quantum computers are significantly resource-constrained and only offer limited and costly access for researchers.
Hence, it is worthwhile to optimize programs prior to execution.
The optimization of pure quantum programs is an active field of research~\cite{karuppasamy2024quantum,fosel2021quantum,ruiz2024quantum,li2024quarl}.
In contrast, to the best of our knowledge, no work exists on the optimization of hybrid quantum programs.

In this work, we take a first step towards optimization of hybrid programs by investigating the following research questions:
\begin{description}
    \item[RQ 1] Which optimizations are applicable to hybrid quantum programs?
    \item[RQ 2] How can the effectiveness of the optimization routines be measured?
\end{description}

To answer these questions, we introduce seven optimization routines and three metrics to evaluate the effectiveness of the optimization.
We implement these routines for the hybrid quantum language Quil~\cite{smith2016practical} and show that our optimizations improve programs according to our metrics.
This lays the foundation for new kinds of hybrid optimizers that enable real-time collaboration between quantum and classical devices.

This work is structured as follows:
We will first give an overview of related work about quantum optimization (cf.~\Cref{sec:related-work}).
Then we will give background information about compiler construction, real-time quantum architecture and the programming language Quil (cf.~\Cref{sec:background}).
We will explain how we analysed Quil programs in \Cref{sect:analysis}.
We will present metrics to evaluate quantum-classical algorithms (cf.~\Cref{sect:metrics}) and propose optimization strategies (cf.~\Cref{sect:optimization-strategies}).
In the end, we will evaluate our optimization strategies in \Cref{sect:evaluation} before we discuss future work (cf.~\Cref{sec:discussion}).

\section{Related Work}\label{sec:related-work}

Optimizing quantum circuits is an active field of research~\cite{karuppasamy2024quantum,fosel2021quantum,ruiz2024quantum,li2024quarl}.
The optimization of quantum circuits involves NP-hard problems, like T-count optimization~\cite{van2023optimising}, a fault-tolerant implementation of topological error-correction~\cite{herr2017optimization}, or parameter optimization with specific requirements on the circuit~\cite{van2024optimal}.
Research about quantum circuit optimizations usually does not consider the interaction between classical and quantum components of a larger computing system~\cite{karuppasamy2024quantum,fosel2021quantum,ruiz2024quantum,li2024quarl}.
This also holds for Quilc~\cite{smith2020open}, the compiler provided for Quil code. This compiler offers optimization routines for quantum circuits only.
To the best of our knowledge, no work exists on the optimization of hybrid quantum programs, and no hardware manufacturer implements any such optimizations.

There are different metrics that quantum circuits have been optimized against.
One of these metrics is the number of gates in a circuit.
An automated optimization has been proposed by Nam et al.~\cite{nam2018automated} that reduced the gate count.
Likewise, Bae et al.~\cite{bae2020quantum} reduced the gate count by creating a quantum mechanical version of Karnaugh maps~\cite{karnaugh1953map}.
Puram, Karuppasamy and Thomas~\cite{puram2024optimizing} lowered the number of gates and the circuit depth by exploiting algebraic expressions representing the quantum circuit.

Some works focus on reducing one type of gate.
For example, Gheorghiu et al.~\cite{gheorghiu2022reducing} proposed a method to reduce the CNOT gate count.
This helps to simplify the mapping of logical to physical qubits in \acp{qpu} with limited connectivity.

Other research uses reinforcement learning in order to optimize quantum circuits.
Fösel et al.~\cite{fosel2021quantum} presented a reinforcement learning approach which showed to be able to reduce the depth and gate count on random 12-qubit circuits.
Nägele and Marquardt~\cite{nagele2024optimizing} applied reinforcement learning on ZX-diagrams~\cite{coecke2008interacting} that represent quantum circuits.
Quietschlich, Burgholzer and Wille~\cite{quietschlich2023compiler} used methods from classical compiler optimization to create a reinforcement learning approach.
Ruiz, Laakkonen and Bausch~\cite{ruiz2024quantum} introduced AlphaTensor-Quantum, a method to optimize the T-count of a circuit.
AlphaTensor-Quantum uses deep reinforcement learning.
Li et al.~\cite{li2024quarl} developed Quarl, a quantum circuit optimizer working with reinforcement learning.
It decomposes the action space into two parts.

Another way to do quantum circuit optimization is pattern matching.
This technique is about finding all matches of a small circuit (pattern) in a large circuit, while considering the commutation relations of quantum gates.
A framework that exploits pattern matching is QCIR, developed by Chen et al.~\cite{chen2022qcir}.
QCIR internally uses a pattern description format, introduced in the same work.
Iten et al.~\cite{iten2022exact} presented a classical algorithm which provably finds all maximal matches for a pattern matching algorithm on a quantum circuit.

Another approach is to use genetic programming to enforce quantum circuit optimization.
Gemeinhardt et al.~\cite{gemeinhardt2025gequpi} proposed \mbox{QeQuPI}, a framework that automatically debugs and improves quantum circuits using genetic programming.
Similarly, Wei et al.~\cite{wei2021genetic} developed an automatic circuit optimization method that is based on genetic programming.

Multiple optimizers have been developed in other research.
Pointing et al.~\cite{pointing2024optimizing} created Quanto, a quantum circuit optimizer that considers circuit identities and automatically generates new ones.
Similarly, the quantum circuit optimizer Quartz by Xu et al.~\cite{xu2022quartz} automatically generates and verifies circuit transformations for arbitrary gate sets.
Paykin et al.~\cite{paykin2023pcoast} presented PCOAST, which does optimizations by exploiting commutation relations of Pauli strings. 
POCAST considers during its optimization procedures whether the quantum state of the qubits needs to be preserved at the end of the calculation, or if we are only interested in measurement outcomes.
Hietala et al.~\cite{hietala2021verified} developed VOQC, a verfied optimizer for quantum circuits. 
It uses the Coq proof assistant~\cite{coq2024} for verification. 
Coq functions are applied to transform the quantum circuit and are proofed to be correct. 

There does exist research that considers the inclusion of quantum computers in classical computing systems~\cite{wille2024qdmi,almudever2024designing,kaya2024software,basermann2025quantum}.
These works, however, only consider pure quantum programs or the construction of a complete software stack for the use of quantum computers.
Our work, in contrast, considers one particular level of such a software stack as will be required for future hybrid computing applications.

\section{Background}
\label{sec:background}

In this section, we provide an overview about classical compiler construction in \Cref{subsect:compiler-construction}, describe the architecture for real-time quantum computation in \Cref{subsect:real-time-quantum-comp} and introduce the Quil language in \Cref{subsect:quil-language}.

\subsection{Compiler Construction}\label{subsect:compiler-construction}

Modern compilers typically go far beyond simply translating a program from a source language into a target language.
Instead, they optimize for the target language as well as the target environment.
Typically, they perform multiple machine-independent as well as machine-dependent passes.
Designing novel optimization passes and choosing the optimal order of passes for a given program remains an active field of research~\cite{ThunigJohannfunkeWangEtAl2024,LiuRenXieEtAl2024,AhmedFahimUlHaqueRazaKhanEtAl2024,ashouri2018survey,triantafyllis2003compiler}.

Each compiler pass either gathers information about the program, or it changes its structure without changing its semantics.
We call the former and latter passes \emph{analyses} and \emph{transformations}, respectively.
Typically, transformations use the information extracted by analyses.

\begin{table}
	\centering
	\caption{Typical analyses and transformations~\cite{aho2007compilers}.}
	\begin{tabular}{ll}
		\toprule
		Analyses & Transformations\\
		\midrule
		Available expressions & Constant folding\\
		Constant propagation & Copy propagation\\
		Live-variable analysis & Dead code elimination\\
		Reaching definitions & \\
		\bottomrule
	\end{tabular}
	\label{table:Analysations_and_transformations}
\end{table}

There exist a plethora of complex analyses and transformations.
In this work, we adapt a number of classical optimization passes to the setting of hybrid programs.
We present these operations in~\Cref{table:Analysations_and_transformations} and provide a brief overview below. 
 For a more detailed discussion, we refer the reader to the work of Aho, Lam, Sethi, and Ullman~\cite{aho2007compilers}.

\textbf{Available expressions} is about checking whether the result of evaluating an expression is available at a program point $p$.
An expression is available if it is evaluated on every possible path between the start of the program and p.
Additionally, there must not be new assignments to $a$ or $b$ between the last evaluation and p.

\textbf{Constant-propagation} means to check whether a variable holds a unique constant value at a program point p.

\textbf{Live-variable analysis} is used to determine which variables are still ``in use'' (alive) at a point p of the program.
If there is any usage of a variable's value between p and the program halt, it is alive.
Otherwise, it is dead.
Values of dead variables have no influence on the remaining program, and they do not need to be considered any further.

\textbf{Reaching definitions} is used to determine which definitions of a variable can reach a program point p.

\textbf{Constant folding} is evaluating constant expressions and replacing the expressions by their values.

\textbf{Copy propagation} checks when a copy statement (e.g. \texttt{a = b}) is given, whether $a$ can be removed and $b$ be used instead of $a$.

\textbf{Dead code elimination} removes ``dead'' (or useless) code from the program.
This affects unreachable code as well as code that computes values which are never used.

\subsection{Real-Time Quantum Architecture}\label{subsect:real-time-quantum-comp}

Real-time communication refers to the exchange of information between the CPU and the \ac{qpu} during the qubit coherence time. 
This communication is crucial for tasks such as applying gates to qubits based on prior measurements or adjusting parameters of quantum gates in response to previous measurement outcomes.
It is mandatory for some algorithms and can be used to reduce quantum resources of a calculation or to check whether a desired quantum state has been created.
Examples for real-time algorithms are quantum teleportation~\cite{bennett1993teleporting}, active reset~\cite{lubinski2022advancing}, magic state distillation~\cite{knill2004fault,bravyi2005universal,fowler2012surface}, repeat-until-success~\cite{paetznick2013repeat,fu2023proto}, iterative phase estimation~\cite{dobvsivcek2007arbitrary,fu2021quingo}, and Shor's algorithm for $2n+1$ qubits~\cite{beauregard2002circuit,rossi2022using}.

\begin{figure}[t]
	\centering
	\includegraphics[height=2.5cm]{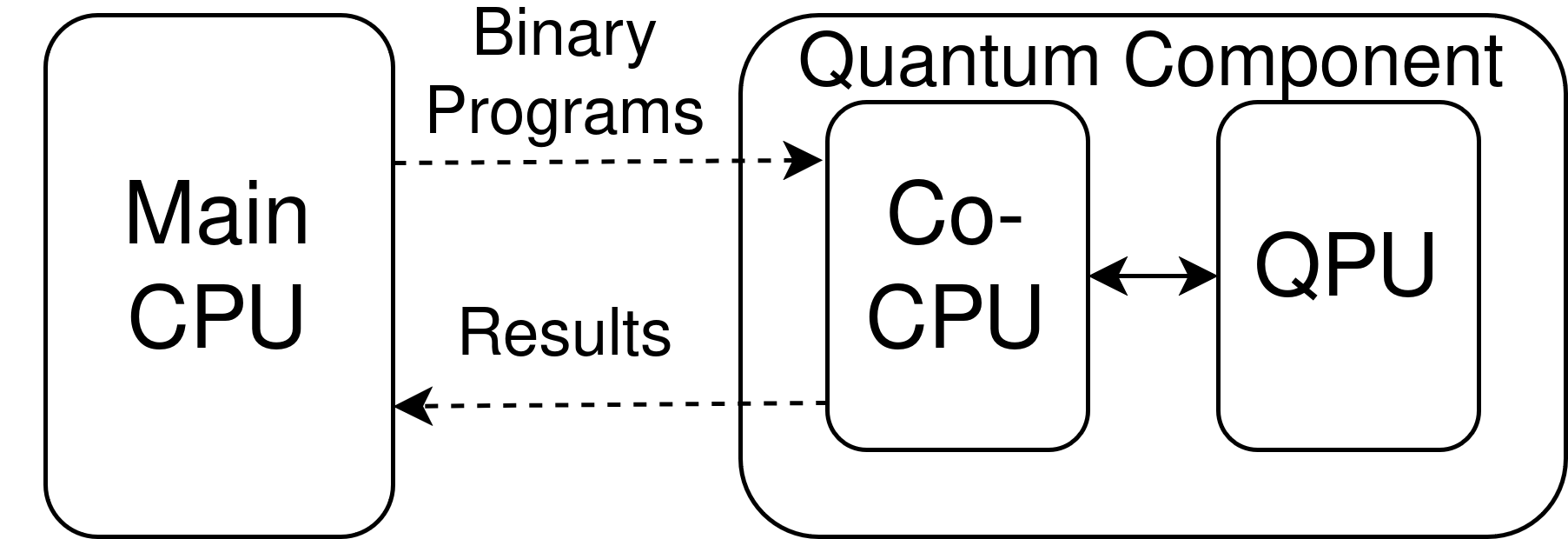}
	\caption{The refined HQCC model (from~\cite[Fig.~2]{fu2021quingo}). Dotted lines indicate slow communication (longer than the coherence time of the qubits), the continuous line indicates fast communication (shorter than qubit coherence time).}%
	\label{fig:hqccRefined}
\end{figure}

To execute real-time quantum computation, a refined \ac{hqcc} architecture is needed~\cite{fu2021quingo} (see \Cref{fig:hqccRefined}).
We will assume this architecture in this work.
A co-CPU and the QPU form a quantum component, which gets a quantum program from the main CPU and returns the results to said main CPU.
This component can apply real-time feedback on qubits and execute classical instructions.
The classical calculations are limited by the coherence time of the qubits, but not by the abilities of the co-CPU.

The \ac{hqcc} architecture provides an interesting hardware case, as we work with two calculation components:
The co-CPU and the QPU.
Both devices work in parallel, and only influence each other on defined points (e.g.\,measurements).

In the following, we will refer to the co-CPU as CPU, unless stated otherwise.

\subsection{The Quil Programming Language}\label{subsect:quil-language}

Quil~\cite{smith2016practical} is an assembly-style low-level language developed by Rigetti to specify quantum-classical computations on an abstract machine architecture.
In this subsection, we give an overview of the instructions that Quil provides, to the extent that it is relevant in this work.
For a more extensive overview, we refer the reader to the paper introducing Quil~\cite{smith2016practical}.

The Quil ecosystem consists of the language specification as well as some additional tools:
PyQuil~\cite{computing2019pyquil}, a Python library to generate and execute Quil code.
Quilc~\cite{smith2020open}, a compiler that compiles Quil circuits to circuits a defined hardware can execute.
And the Quil-Lang \ac{qvm}~\cite{qvm_quil_project}, that can simulate the execution of Quil code.

In Quil, integer indices are used to refer to qubits.
Quil supports four classical variable types: \texttt{BIT}, \texttt{OCTET}, \texttt{INTEGER}, and \texttt{REAL}.
Classical variables need to be declared before usage, e.g. \texttt{DECLARE~a~BIT}.
Qubits do not need to be declared upfront.

Quil offers different categories of instructions, namely
\begin{itemize}
	\item \textbf{quantum gates}: Instructions that name a gate and at least one qubit that is applied to the gate, e.g.\,\texttt{H~0}, or \texttt{CNOT~0~1}.
    \item \textbf{parameterized quantum gates}: Quantum gate instructions with gates that receive a classical parameter, e.g.\,\texttt{RZ(angle)~0}, with the classical parameter \texttt{angle}. Quantum gates that receive a fixed value (e.g.\,\texttt{RZ(1.57)}) do not count as parameterized in the context of this work.
	\item \textbf{classical instructions}: Calculations applied on classical variables. The classical instruction set is turing-complete. It is comparable to the instruction set of the Assembly language~\cite{assemblyManual}: For example, \texttt{MOVE} writes the value of the second variable into the first, \texttt{ADD} adds the value of the second variable onto the value of the first, and \texttt{NEG} negates the given variable~\cite[6.5]{quil_language_specification}.
	\item \textbf{measurements}: Instructions that measure the value of a qubit. The value can, but does not have to be, saved in a classical parameter. E.g.: \texttt{MEASURE~0~ro} or \texttt{MEASURE~1}. 
	\item \textbf{control structures}: One can define labels (\texttt{LABEL~@label\_name}) and jumps. Jumps can be conditional (e.g.\,\texttt{JUMP-WHEN~@label\_name~cond}) or unconditional (e.g.\,\texttt{JUMP~@label\_name}).
\end{itemize}

Quantum gates that are not parameterized can be exclusively executed by a QPU and are therefore \emph{quantum instructions}.
Classical instructions can be exclusively executed by a CPU and are therefore \emph{classical instructions}.
All other instructions, including control structures, need to be executed by the QPU and the CPU at the same time, and are therefore \emph{hybrid instructions}.

We explore the optimization Quil code with respect to heterogeneous architecture.

\section{Analysis of Quil Programs}\label{sect:analysis}

In order to optimize Quil programs, we first have to analyze their structure.
We will optimizie the analyzed Quil programs in \Cref{sect:evaluation}.

To the best of our knowledge, no benchmark database for the real-time quantum-classical use case exists. 
As most works focus on quantum circuits, there are relatively few algorithms with real-time feedback properties. 
We chose to work with four algorithms that are mentioned frequently in literature. 
Creating an extended benchmark database is out of scope for our work and is a topic for future work.

The four real-time quantum classical algorithms used in this paper are:
\begin{itemize}
	\item Quantum teleportation~\cite{bennett1993teleporting}
	\item Magic state distillation~\cite{knill2004fault,bravyi2005universal,fowler2012surface} 
	\item Repeat-until-success~\cite{paetznick2013repeat,fu2023proto}
	\item \ac{ipe}~\cite{dobvsivcek2007arbitrary,fu2021quingo}
\end{itemize}

The code for the algorithms is given in~\cite{supplementary_github}, alongside the results of the analysations described in this section.

In this section, we will describe how to do control flow analysis and data-dependence analysis on Quil.

\subsection{Control Flow Analysis}

\begin{figure}[t]
	\centering
	\includegraphics[width=.48\textwidth]{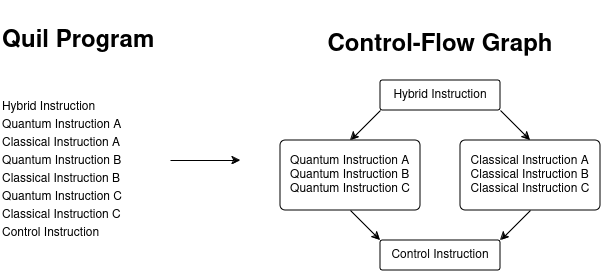}
	\caption{Creating basic blocks in the CFG from alternating quantum and classical instructions.}%
	\label{fig:cfg-divison}
\end{figure}

A way of analysing a program for different compilation steps is to do a control flow analysis~\cite{allen1970control}.
This can be done using a \ac{cfg}.
A \ac{cfg} represents all possible execution paths of a program.
The directed edges of the graphs represent jumps.
The nodes describe a basic block, which are ``sequences of statements that are always executed one-after-the-other, with no branching''~\cite[chapter~2.8.1]{aho2007compilers}.
This means that a jump on the current instruction as well as a label (jump target) at the next instruction end a basis block~\cite[chapter~8.4.1]{aho2007compilers}.

We created a \ac{cfg}~\cite{allen1970control} for all programs we evaluate.
To show which device executes which instructions, we divide the instructions depending on their type.
Each node/basic block of the \ac{cfg} consists of either only quantum instructions, only classical instructions, or only hybrid/control structures.
Quantum and classical instructions that are executed without a hybrid/control instruction in between are written into two parallel basic blocks.
An example of this is shown in~\Cref{fig:cfg-divison}.
By creating a \ac{cfg} in this way, one can easily see which parts of the code are executed by which device(s).

\subsection{Data-Dependence Analysis}

Most optimization techniques depend on data-flow analysis~\cite[chapter~9.2]{aho2007compilers}.
One way to do data-flow analysis is to create a \ac{ddg}\footnote{Also called program-dependence graph~\cite[chapter~11.8.2]{aho2007compilers}.}~\cite[chapter~5.3.2]{cooper2022engineering} of the program.
A \ac{ddg} is a directed graph.
Nodes represent basic blocks (or, in our case, single instructions) of a program.
An instruction $A$ is a successor of instruction $B$ in the graph, if $A$ has to be executed after $B$ in order for the program to be correct, e.g. if $A$ and $B$ access the same variable.
Edges only appear between instructions/nodes if no other instruction $C$ has to be executed between $A$ and $B$.

We create a \ac{ddg} using information from the Quil program and the \ac{cfg}.
Every node of the \ac{ddg} holds a single Quil instruction.
The edges indicate the order in which the instructions have to be executed.
This is done by checking variable dependency.
Unconditional jumps are resolved before creating the \ac{ddg} and not listed as nodes.

Conditional jumps in the program pose a problem to the \ac{ddg}.
If a conditional jump targets an already executed line, it introduces circular dependencies.
This could only be resolved exactly if we knew the number of iterations, which would be analogous to solving the Halting problem, and therefore not generally possible.

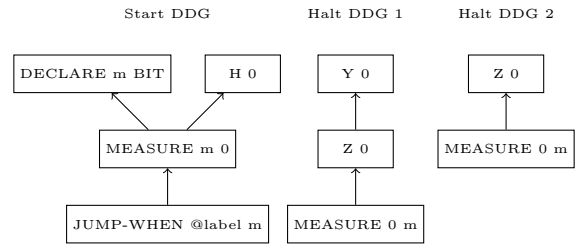
\begin{figure}
	\centering
	\begin{tikzpicture}
    \tiny
	\tikzstyle{block} = [rectangle, minimum width=1cm, minimum height=.5cm, draw=black]
	
	\node (start) [align=left] {Start DDG};
	\node (first) [block, align=left, below of=start, xshift=-1cm, yshift=.2cm] {DECLARE m BIT};
	\node (second) [block, align=left, right of=first, xshift=1cm] {H 0};
	\node (third) [block, align=left, below of=first, xshift=1cm] {MEASURE m 0};
	\node (fourth) [block, align=left, below of=third] {JUMP-WHEN @label m};
	
	\draw [->] (third) -- (first);
	\draw [->] (third) -- (second);
	\draw [->] (fourth) -- (third);
	
	\node (halt1) [align=left, right of=start, xshift=1.5cm] {Halt DDG 1};
	\node (firsth1) [block, align=left, below of=halt1, yshift=.2cm] {Y 0};
	\node (secondh1) [block, align=left, below of=firsth1] {Z 0};
	\node (thirdh1) [block, align=left, below of=secondh1] {MEASURE 0 m};
	
	\draw [->] (secondh1) -- (firsth1);
	\draw [->] (thirdh1) -- (secondh1);
	
	\node (halt2) [align=left, right of=halt1, xshift=1cm] {Halt DDG 2};
	\node (firsth2) [block, align=left, below of=halt2, yshift=.2cm] {Z 0};
	\node (secondh2) [block, align=left, below of=firsth2] {MEASURE 0 m};
	
	\draw [->] (secondh2) -- (firsth2);
	\end{tikzpicture}
	\caption{An example for multiple DDGs originating from one Quil code. The code can be found in \Cref{lst:exampleDDGCode}.}
	\label{fig:multipleDDGs}
\end{figure}

We resolve this issue by creating a \ac{ddg} only up to the next conditional jump.
By that, we can receive multiple \acp{ddg} for a single program, all depicting a part of the program.
One of the \acp{ddg} is the start \ac{ddg}, which is the \ac{ddg} describing the entry of the program.
Additionally, we have one or multiple halt \acp{ddg}, which include the program instructions last executed before the program terminates.
All DDGs beside the start DDG start at the jump target of a conditional jump. 
If an instruction is the target of multiple conditional jumps, we create multiple DDGs.
An example for a program that results in multiple \acp{ddg} is given in \Cref{lst:exampleDDGCode}, and the corresponding \acp{ddg} in \Cref{fig:multipleDDGs}.

\begin{lstlisting}[caption={An example for Quil code that results into multiple DDGs. The corresponding DDGs can be found in \Cref{fig:multipleDDGs}.},label={lst:exampleDDGCode}]
DECLARE m BIT
H 0
MEASURE 0 m
JUMP-WHEN @label m
Y 0
LABEL @label
Z 0
MEASURE 0 m		
\end{lstlisting}

For the \ac{ipe}, we resolved conditional jumps before \ac{ddg} creation.
In the executable \ac{ipe}, we use the code given in \Cref{lst:executableAdding} to add $1$ to \texttt{param\_no\_pi[0]} if \texttt{lastMeasurement[0]} is $1$.

\begin{minipage}{\linewidth}
\begin{lstlisting}[caption={Executable version of adding $1$ to \texttt{param\_no\_pi[0]} if \texttt{lastMeasurement[0]} is true. Quil's semantic does not allow to add a \texttt{BIT} variable to a \texttt{REAL} variable. Therefore, we need conditional jumps.},label={lst:executableAdding}]
DECLARE param_no_pi REAL[1]
DECLARE lastMeasurement BIT[1]

JUMP-UNLESS @noadd1 lastMeasurement[0]
ADD param_no_pi[0] 1
LABEL @noadd1
\end{lstlisting}
\end{minipage}

Conditional jumps have to be used, as Quil's semantic does not allow adding a bit-value to a real value.
In the \ac{ipe} that is used for the \ac{ddg} and during the optimization, the above code is changed to the one in \Cref{lst:analyzationAdding}:
The bit value is directly added to the real value.

\begin{lstlisting}[caption={Version we use for analysation of adding $1$ to \texttt{param\_no\_pi[0]} if \texttt{lastMeasurement[0]} is true. The \texttt{BIT} value is directly added to \texttt{param\_no\_pi[0]} to avoid conditional jumps.},label={lst:analyzationAdding}]
DECLARE param_no_pi REAL[1]
DECLARE lastMeasurement BIT[1]

ADD param_no_pi[0] lastMeasurement[0]
\end{lstlisting}

While Quil's semantic forbids this construction, it has the same logical effect:
Adding~$1$ to \texttt{param\_no\_pi[0]} if \texttt{lastMeasurement[0]} is $1$.
This is done to prevent conditional jumps in the \ac{ipe} algorithm, which makes the optimization and its evaluation simpler.
This has only been done for \ac{ipe}, not for the other algorithms.
The \acp{cfg} and \acp{ddg} for all algorithms can be found in our GitHub repository alongside our code~\cite{supplementary_github}.

\section{Metrics for Quil Programs}\label{sect:metrics}

How to best evaluate quantum-classical algorithms is an open question.
While there are some metrics for the evaluation of quantum circuits available (cf.~\Cref{sec:related-work}), extending them to the quantum-classical case is not straightforward. 
Two widely used metrics, the numbers of gates and the circuit depth, do not take into account classical calculation nor interaction between classical and quantum calculations.

In this work, we assume a relatively simple execution model, which has an execution time of 1 per instruction and no communication latencies.
This model does not accurately depict real-world quantum computers, but suffices for our proof of concept.
A more accurate description is dependent on the hardware and out of scope for this work.
We plan to access this question in future work.

We propose three metrics to statically evaluate quantum-classical Quil programs: Wall time, \ac{qin}, and \ac{qct}.
The metrics can be used to optimize Quil programs against or to evaluate how well optimization methods worked.
Our GitHub repository~\cite{supplementary_github} provides the functionality to evaluate Quil programs with respect to our metrics.
We evaluate our optimization methods against our metrics.
The results are given in \Cref{sect:optimization-strategies}.

\subsection{Wall Time}

The wall time of a program is the total time from the start to the end of a program.
To calculate the wall time of a hybrid quantum-classical algorithm, we need to consider that the CPU and the QPU work in parallel.

\begin{figure}
	\centering
	\includegraphics[width=.48\textwidth]{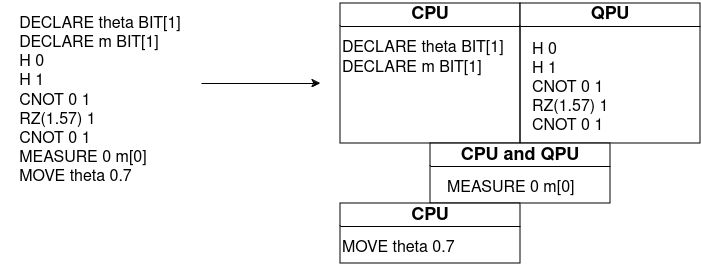}
	\caption{An example of a Quil execution that has a wall time of $7$.}%
	\label{fig:naive-quil-execution}
\end{figure}

We asses how long the program would take to execute while considering that all instructions have an execution time of $1$. 
Whenever possible, the CPU and the QPU execute instructions in parallel. 
Two parallel executed instructions receive an execution time of $1$ together.
At a hybrid or control instruction, one device waits for the other. 
For example, the program in \Cref{fig:naive-quil-execution} has a wall time of $7$.
	
The wall time is calculated for all \acp{ddg} of a program and summed up for comparison of syntactically different representations of the same program.
This means that if two identical DDGs exist due to instructions being targets of multiple jumps (cf.~\Cref{sec:related-work}), that DDG is counted twice.

\subsection{Quantum Instruction Number}

Quantum instructions can introduce error into a calculation, as quantum gates of \ac{nisq} devices do not have perfect fidelity.
Therefore, minimizing the \ac{qin} is sensible.

The number of quantum instructions per \ac{ddg} are counted and summed up for comparison to other codes.
Hybrid instructions are included in the counting, as the QPU and the quantum state are considered during hybrid calculations.
	
\subsection{Quantum Calculation Time} 

As quantum states in a QPU have a limited coherence time, it is sensible to minimize the time the QPU has to keep the qubits coherent,\,i.e. the time during which the QPU calculates.
This value is the \ac{qct}.
As we assume that each instruction needs a time value of $1$, the time is given in units of the instruction number.

For the calculation of the \ac{qct}, we assume the best case:
The QPU starts with the first quantum instruction on the latest possible time such that the CPU does not have to wait for the QPU at the first hybrid instruction.
Further, the QPU calculates the quantum instructions after the last hybrid instruction directly after the hybrid instruction.
Additionally, the QPU has to keep the quantum state stable during the wall time between the first and last hybrid instruction.

It should be noted that a reasonable program has no quantum instructions after the last hybrid instruction.
This would only create quantum information which is destroyed again, as it would have to be measured for further usage.
Measurement, again, would be a hybrid instruction.

This results in a \ac{qct} $\tau_Q$ of
\begin{align}
	\tau_Q &= \delta t_\mathrm{between} + n_\mathrm{q\_before} + n_{\mathrm{q\_after}}
\end{align}
with the wall time between first and last hybrid instruction (including these hybrid instructions) $\delta t_\mathrm{between}$, the number of quantum instructions before the first hybrid instruction $n_\mathrm{q\_before}$, and the number of hybrid instruction after the last hybrid instruction $n_\mathrm{q\_after}$.
An example can be seen in \Cref{fig:qet-calculation}.

\begin{figure}
	\centering
	\begin{tikzpicture}
	   \node (code) [align=left] {DECLARE r REAL\\DECLARE m BIT\\H 0\\Y 0\\MEASURE 0 m\\H 0\\Z 0\\MOVE r 2\\RZ(r) 0\\H 0\\MEASURE 0 m\\Z 0};
	   \node (tyoes) [align=left,right of=code, xshift=2cm] {Classical\\Classical\\Quantum\\Quantum\\Hybrid (first)\\Quantum\\Quantum\\Classical\\Hybrid\\Classical\\Hybrid (last)\\Quantum};
	   \draw [decorate, decoration = {calligraphic brace}] (-2,0.9) -- node[left=6pt] {$n_\mathrm{q\_before} = 2$}  (-2,2.5);
	   \draw [decorate, decoration = {calligraphic brace}] (-2,-2.0) -- node[left=6pt] {$\delta t_\mathrm{between} = 6$}  (-2,0.8);
	   \draw [decorate, decoration = {calligraphic brace}] (-2,-2.5) -- node[left=6pt] {$n_\mathrm{q\_after} = 1$}  (-2,-2.1);
	\end{tikzpicture}
	\caption{An example of determining $\delta t_{between}$, $n_{q\_before}$ and $n_{q\_after}$ from Quil code. The QCT is the sum of these three values.}
	\label{fig:qet-calculation}
\end{figure}
	
A difficulty of this metric arises when the program is divided into more than one \ac{ddg}.
The first hybrid instruction in the start \ac{ddg} is the globally first hybrid instruction.
The last hybrid instruction in the halt \ac{ddg} is the globally last hybrid instruction.
If there are multiple possible last \acp{ddg}, we take the one that causes the longest \ac{qct} (worst-case).
The \ac{qct} is calculated by the number of quantum instructions before the first hybrid instruction and after the last hybrid instruction.
Additionally, the wall time of all other \acp{ddg} and between the first and the last hybrid instruction (included) are added.
	
If the last \ac{ddg} has no quantum instruction, it does not add to the QCT.
The wall time of all other \acp{ddg} is simply added, as every \ac{ddg} necessarily ends at a hybrid conditional jump.

Note that, in an ideally optimized program, a \ac{ddg} with only classical instructions should not exist, as this could be calculated completely by a main CPU and does not have to be sent to a CPU-QPU complex.

\begin{table}
	\centering
	\caption{Instruction Number (instr. no.), wall time, QIN, and QCT of the original Quil files of the different algorithms. The instruction number and wall time are calculated per DDG.}
	\begin{tabularx}{.48\textwidth}{ccccc}
		\toprule
		Example program & Instr. no. & Wall time & QIN & QCT\\
		\midrule
		Quantum teleportation & $8$, $2$, $2$ & $6$, $2$, $1$ & $9$ & $10$  \\  
		Magic state distillation & $67$, $63$, $6$ & $62$, $62$, $6$ & $66$ & $130$ \\
		Repeat-until-success & $12$, $11$, $10$, $7$ & $9$, $10$, $9$, $6$ & $34$ & $35$ \\  
		IPE & $55$ & $45$ & $25$ & $33$ \\
		\bottomrule
	\end{tabularx}
	\label{table:Initial_properties}
\end{table}

\Cref{table:Initial_properties} shows the values of the different metrics for our initial implementations of the algorithms.
We will look at strategies to improve (i.e.,\,reduce) the values in the next section.

\section{Optimization Strategies}\label{sect:optimization-strategies}

In this section, we aim to create a set of optimization methods for heterogeneous quantum-classical architectures.
The optimizations are done statically by a main CPU, before the program is send to the CPU-QPU component.
Our GitHub repository~\cite{supplementary_github} provides the functionality to apply these optimization methods to a Quil code.

In the beginning, we will derive some optimization strategies from already well established classical (machine-independent) strategies.
Afterwards, we will design strategies specifically targeting the quantum-classical nature of real-time code.

\subsection{Strategies Adapted from the Classical Case}

We will examine if and how the classical machine-independent optimizations mentioned in \Cref{subsect:compiler-construction} can be applied to the quantum-classical heterogeneous case.
Some optimization methods are not sensible for the quantum case, but could in principle be implemented for the classical part of the algorithms.
We did not implement these optimizations, to keep our focus on quantum-classical hybrid code.

It is insensible to try to determine \textbf{available expressions} in Quil.
Instructions that calculate an expression $a \bullet b$ and assign the result to another variable do not exist.
Instead, quantum operators act directly on qubits.
Additionally, the results of classical calculations are stored in one of the involved variables.
For example, 
\begin{align}
    \texttt{ADD a b}
\end{align}
is interpreted as
\begin{align}
    a \coloneqq a + b.
\end{align}

We apply \textbf{constant propagation} to the classical and the quantum part of the code.

For the classical part of the code, the algorithm works as follows:
A classical value is recognized as constant by a \texttt{MOVE} instruction if the instruction moves a constant value onto it.
E.g 
\begin{align}
	\texttt{MOVE a 10}
\end{align}
 results into \texttt{a} having the constant value $10$.

A variable remains constant until it is being written to again.
\begin{align}
	\texttt{ADD b a}
\end{align}
writes and reads \texttt{b}, but only reads \texttt{a}, thus \texttt{a} remains constant.
This means if \texttt{b} was constant before the instruction the algorithms recognizes \texttt{b} to be constant at the reading part of the instruction, but \texttt{b} looses its constant state afterward.
This would even hold if \texttt{a} was replaced by a constant value.
Constant folding will replace the instruction by a \texttt{MOVE} instruction if \texttt{a} and \texttt{b} were constant at read-time.

For the quantum part of the code, we restrict ourselves to the Pauli-basis and single qubit gates.
This was done to avoid an exponential scaling of the constant propagation algorithm while trying to keep track of arbitrary quantum states.

A qubit is constant with the value \texttt{Pauli~X~Zero} at the start of the program and after a \texttt{RESET}.
The constant propagation checks at which instructions a constant qubit ``arrives'' with a specific value, either because it is initially used or has been reset before, or because constant folding has shown that it remained in a constant Pauli basis after a gate application.

\textbf{Live-variable analysis} can be used for the classical and quantum parts of heterogeneous programs.

\begin{lstlisting}[caption={An example of Quil code with live and dead variables. Assuming \texttt{a} is a readout variable, it counts as dead in line $4$, as the addition result is not read anywhere before a is written to in line $6$. \texttt{b} is dead in line $5$, as it is not read afterward.},label={lst:liveVarClassicalQuil}]
DECLARE a INTEGER
DECLARE b INTEGER
MOVE a 3
ADD a 10
MOVE b 7
MOVE a 10
\end{lstlisting}

A classical variable counts as dead if it is not read until the variable's value is overwritten or the program terminates.
For the application of this algorithm, we need to know which classical variables hold \emph{readout values}, i.e.\,which values need to be reported back to the main CPU at the end of the quantum-classical program.
These variables are by default alive at the end of the program, which is handled as if there was an additional read of the variable after the halt of the program.
An example is given in \Cref{lst:liveVarClassicalQuil}.

A qubit counts as dead if its information content does not influence any classical information from the current instruction until the end of the program.

The qubit can safely be called dead at a program point $p$ if it is
\begin{itemize}
	\item not used by any multi-qubit gate, \textbf{and}
	\item not measured with the result saved in a classical variable
\end{itemize}
until
\begin{itemize}
	\item the end of the program, \textbf{or}
	\item the next reset.
\end{itemize}

The variable may be entangled with other variables.
Due to the no-communication theorem~\cite{peres2004quantum}, operations on a qubit $A$ cannot influence the measurement result on other qubits without using a measurement result of $A$.
This even holds if $A$ and the other qubits are entangled.
Therefore, we do not have to check for entanglement if a qubit will not be used by any multi-qubit gates in succeeding instructions and if the result of a measurement on it will not be considered in the program.

We can only apply live-variable analysis at the halt \ac{ddg} of the program, as we cannot say whether a variable will be used in future \acp{ddg}.
This limits the usage of the live-variable analysis.
Live-variable analysis could be applied to non-halt \acp{ddg} in a restricted manner:
All values count as alive at the end of the \ac{ddg} (handled as if a read-instruction is about to follow).
If a variable is dead because its value is overwritten within one \ac{ddg}, it can still safely be called dead.
However, our current implementation~\cite{supplementary_github} does not support this.

\textbf{Reaching definitions} can be implemented for the classical Quil variables by checking \texttt{DECLARE} and \texttt{MOVE} statements.
Quil programs do not offer define/declare statements for qubits, neither direct assignment of a value to a qubit.
Therefore, applying reaching definitions to the quantum part of the program is not sensible.
As we focused on quantum-classical routines, we did not implement reaching definitions for the classical part of the programs either.

\textbf{Constant folding} can be applied on classical and quantum parts of our hybrid code.
In classical instructions, all constant variables that are read are replaced by their constant values. 
This means, e.g.,
\begin{align}
	\texttt{ADD a b}
\end{align}
is replaced by
\begin{align}
	\texttt{ADD a 10}
\end{align}
if \texttt{b} has a constant value of $10$.
The same holds for parameterized quantum gates. If the classical parameter is constant, it is replaced by the constant value.

If a value that is written to is constant and additionally, all read values in the instructions are constant, the instruction is replaced by a \texttt{MOVE} instruction.
As an example: If \texttt{a} is $5$, \texttt{b} is $7$ and the instruction is
\begin{align}
	\texttt{ADD a b},
\end{align}
it is replaced by
\begin{align}
	\texttt{MOVE a 12}
\end{align}
as \texttt{a} would be $12$ after the addition.

For the quantum part of the program, constant folding calculates the result of the application of a gate on a qubit.
This is only calculated if the qubit is constant beforehand (i.e.\,in a Pauli basis, as defined by constant propagation) and a single-qubit Clifford gate acts on it.
We calculate the next Pauli state and still assume the qubit as constant (i.e.\,in the respective Pauli state).

Constant folding for qubits can be done efficiently for Clifford gates due to the Gottesman-Knill theorem~\cite{gottesman1998theory}.
To avoid dealing with entangled states during constant propagation, we generally assume qubits to be non-constant after a multi-qubit gate.

Due to this, we only trace the constant value of a qubit until the first non single-qubit or non-Clifford gate acts on it.
This means we are tracing the six Pauli basis states as values of the qubits.

Measurements are in general considered to be random and write non-constant values in the classical values.
The only exception is if we know that a qubit is either in the $\ket{0}$ or $\ket{1}$ state.
In that case, the classical value receives the constant value $0$ or $1$.

\textbf{Copy propagation} can be used in the classical part of heterogeneous code.
It is not sensible for quantum code due to the no-cloning theorem~\cite{wootters1982single,dieks1982communication}. 

\textbf{Dead code elimination} can be applied to classical and quantum instructions.
We use the results of the live-variable analysis about which variables are dead.
Classical code is dead if all variables written to in the instruction are dead.
Quantum code is dead if all qubits an instruction acts on are dead.
This also holds for parametrized quantum gates.
All dead instructions are deleted.

These were optimization methods taken from the compilation routine of purely classical code.
We end up with \emph{constant propagation}, \emph{live-variable analysis}, \emph{constant folding}, and \emph{dead code elimination} for the quantum-classical case.

\subsection{Strategies Designed for Quantum-Classical Calculations}

Besides the classical methods, we can apply algorithms that are specific for the heterogeneous quantum-classical architecture.
We introduce three algorithms targeting quantum-classical architecture in this work:
An analysis operation that finds hybrid-dependencies, and the two transformation operations instruction reordering and latest possible quantum execution.

\textbf{Finding hybrid-dependencies} checks which instructions have to be executed before a hybrid node becomes executable.
This analysis first finds all hybrid instructions, and afterwards determines which instructions have to be executed before the respective hybrid instruction by using the \ac{ddg}.
A hybrid instruction can have other hybrid instructions as dependencies.
A hybrid instruction ``blocks'' the instruction dependencies for all other hybrid instructions.

\begin{lstlisting}[caption={An example of Quil code with hybrid instructions that depend on previous lines.},label={lst:hybridDependenciesQuil}]
DECLARE m INTEGER[1]
H 0
MEASURE 0 m
RZ(m) 0
\end{lstlisting}

For example, \Cref{lst:hybridDependenciesQuil} contains the hybrid instructions \texttt{MEASURE} and \texttt{RZ(m)}.
All other lines have to be executed before \texttt{RZ(m)}.
But the only direct dependency we save for it is \texttt{MEASURE}, as \texttt{MEASURE} is a hybrid instruction and depends on \texttt{H} and the \texttt{DECLARE} instruction as well.
For \texttt{MEASURE}, the dependencies \texttt{H} and \texttt{DECLARE} are saved.

\textbf{Instruction reordering} can be done as long as instructions dependent on each other are still in the same order.
The \ac{ddg} provides information about the dependent instructions.

The goal of this routine is to reorder the instructions to maximize the number of parallel executions of the CPU and QPU.
In other words, the time in which only one device is calculating and the other one is idling should be minimized. 

The two devices only influence each other through hybrid nodes.
To prevent long waiting times of one device for the other, we use the knowledge we gained from the \ac{ddg} and the finding hybrid-dependency analysis.

\begin{figure*}
\begin{lstlisting}[caption={Instruction reordering algorithm.},label={lst:instruction-reordering},mathescape=true]
Input: 
instruction_ddg: The instructions of the programm with the dependencies of the instructions.

Output: 
execution_queue: The (probably new) order in which the instructions should be executed. Must keep the topological order of instruction_ddg.

execution_queue $\leftarrow$ empty list
relevant_instructions_list $\leftarrow$ hybrid instructions and last instruction of instruction_ddg
next relevant_instruction to be executed:
	dependencies $\leftarrow$ instructions that still need to be executed before relevant_instruction in instruction_ddg
	quantum_number $\leftarrow$ amount of quantum instructions in dependencies
	classical_number $\leftarrow$ amount of classical instructions in dependencies
	Add dependencies to execution_queue
	while new instructions in instruction_ddg are excutable wrt current execution queue and quantum_number != classical_number:
		if quantum_number > classical_number:
			Add executable classical instructions to execution queue
		else:
			Add executable quantum instructions to execution queue
		Update quantum_number and classical_number according to the number of added instructions
	add relevant_instruction to execution_queue
	repeat
\end{lstlisting}
\end{figure*}

\Cref{lst:instruction-reordering} shows the pseudocode of the instruction reordering algorithm.
If the CPU or the QPU have to wait for the other device before a hybrid instruction, the algorithm aims to avoid that the waiting device does not execute instructions.
It is checked whether there are instructions for the waiting device that can already be executed, i.e.\,for which all dependencies have been executed.
If this is the case, the waiting device can execute the instructions while waiting instead of idling.

\textbf{Latest possible quantum execution} is one method to keep the total execution time of the quantum device small.

The goal of this algorithm is to execute as many classical instructions as possible before the QPU starts to work.
For this, the execution of instructions is reordered in the following way:
\begin{itemize}
	\item All classical instructions that are not dependent on a quantum instruction are executed at the start.
	\item The first quantum instruction is executed such that the QPU and the CPU reach the first hybrid node simultaneously. This is again done with the assumption that each instruction has an execution time of $1$.
\end{itemize}

After setting up this set of optimization operations, we will check in the next section how these operations could optimize example algorithms.

\section{Evaluation}\label{sect:evaluation}

In this section, we will examine how the optimization operations introduced in \Cref{sect:optimization-strategies} affect the quantum algorithms mentioned in \Cref{sect:analysis}.
Some of the algorithms from \Cref{sect:analysis} are not sensible for the quantum case, but can in principle be applied to the classical part of a quantum-classical algorithm.
To keep our focus on the hybrid code, we only implemented the optimization routines that can be applied on quantum and classical parts of the code.
This results in the implementation of \emph{constant propagation}, \emph{live-variable analysis}, \emph{constant folding}, \emph{dead code eliminiation}, \emph{finding hybrid-dependencies}, \emph{instruction reordering} and \emph{latest possible quantum reordering}.
We implemented these optimization strategies, the codebase can be accessed at our repository~\cite{supplementary_github}.

Determining the best order to apply optimization operations on a given code (phase-ordering) is an open question in classical compiler architecture~\cite{ashouri2018survey,triantafyllis2003compiler}.
Finding a sensible order is out of scope for this work, and we will simply draw operations to apply to the codes at random.
We do this $500$ times for each algorithm and apply $50$ optimization operations every time.
Although the sequence of operations was random, specific analysis and transformation routines had to follow one another in a prescribed order, as certain analysis operations provide the necessary background information for subsequent transformation operations.
The optimization routine consisted of a permutation of the following operations succeeding each other:
\begin{itemize}
	\item Constant propagation $\rightarrow$ constant folding.
	\item Live-variable analysis $\rightarrow$ dead code elimination.
	\item Finding hybrid-dependencies $\rightarrow$ instruction reordering.
	\item Finding hybrid-dependencies $\rightarrow$ latest possible quantum reordering.
\end{itemize}
Which means we did $25$ random draws per optimization routine.

\begin{table}
	\centering
	\caption{Best (i.e. minimal) wall time, instruction number, and QCT that could be reached by random application of $50$ optimizations. No given number indicates that no improvement could be done compared to the original program. The QIN has never been improved. The original values can be found in \Cref{table:Initial_properties}.}
	\begin{tabularx}{.42\textwidth}{cccc}
		\toprule
		Example program & Wall time~ & Instr. no.~ & QCT\\
		\midrule
		Quantum teleportation & -- & -- & -- \\  
		Magic state distillation & $53$, $53$, $6$ & -- & $112$ \\
		Repeat-until-success & -- & -- & -- \\  
		IPE & $35$ & $51$ & $30$ \\
		\bottomrule
	\end{tabularx}
	\label{table:Best_properties}
\end{table}

After all optimization operations had been applied, the metrics of \Cref{sect:metrics} were calculated for the optimized Quil program.
The best resulting values for the metrics to evaluate Quil programs against can be seen in \Cref{table:Best_properties}.

No improvements were observed in comparison to the original programs for quantum teleportation and repeat-until-success protocols.
The \ac{ipe}'s wall time, instruction number, and \ac{qct} could be reduced by $22\%$ (wall time), $7.3\%$ (instruction number), and $9.1\%$ (\ac{qct}).

The magic state distillation's wall time (summed over all \acp{ddg}) could be reduced by $14\%$, its QCT dropped by $14\%$ as well.
It yields the same values for all metrics in every optimization iteration.

\begin{table}
	\centering
	\caption{The results of optimizing the iterative phase estimation with respect to the metrics we evaluate the code against. $500$ runs have been done. The frequencies of the different combinations of wall time, instruction number, QIN and QCT are given. The bold results are the best ones for this metric (only given if the metric does not have the same result for all sets).}
	\begin{tabularx}{.42\textwidth}{ccccc}
		\toprule
		Wall time~ & Instr. no.~ & QIN~ & QCT~ & Result frequency\\
		\midrule
		\bm{$35$} & \bm{$51$} & $25$ & \bm{$30$} & $49.2\%$ \\
		$36$ & \bm{$51$} & $25$ & \bm{$30$} & $43.0\%$ \\
		$36$ & \bm{$51$} & $25$ & $31$ & $6.6\%$ \\
		$37$ & $52$ & $25$ & \bm{$30$} & $0.6\%$ \\
		$37$ & $53$ & $25$ & $31$ & $0.2\%$ \\
		$39$ & \bm{$51$} & $25$ & $33$ & $0.2\%$ \\
		$39$ & $55$ & $25$ & $32$ & $0.2\%$ \\
		\bottomrule
	\end{tabularx}
	\label{table:ipe_results}
\end{table}

We received varying  results for the application of optimization routines on the \ac{ipe} algorithm, as outlined in \Cref{table:ipe_results}.
The most frequent results are a wall time of $35$, an instruction number of $51$, and a QCT of $30$.
This is not only the most frequent result set, but also the result set with most optimal values for all metrics.
The second most frequent result has a value of $36$ for the wall time.

\section{Discussion and Future Work}\label{sec:discussion}

In this work, we demonstrated that it is possible to optimize quantum-classical code based on the metrics we introduced.
While the programs we applied this approach to showed modest improvements, the results highlight the potential for further enhancement in future applications.
As our results indicate, no improvements were observed for quantum teleportation or repeat-until-success. This is likely due to the relatively low number of instructions within these applications, which may limit the potential for optimization.

Nevertheless, we expect the optimization of heterogeneous quantum-classical programs to become more relevant in the future.
The development of more abstract quantum programming languages will lead to more compilation steps between the written program and the hardware, which will also lead to a bigger need of optimization routines.
Therefore, we emphasize that the optimization of quantum-classical code is something that should be investigated in more detail.

For evaluation purposes, we had to generate multiple \acp{ddg} of a Quil program, eliminating circular dependencies and loops to apply our optimization metrics (cf.~\Cref{sect:evaluation}).
In contrast, loops are a significant portion of classical programs, making loop optimization a key factor in improving overall program performance~\cite{aho2007compilers}.

It would be valuable to explore how loop optimizations and branch prediction techniques can be applied to quantum-classical programs. 
One straightforward approach could be to eliminate a conditional jump if static analysis confirms that the result is certain.

For our evaluations we assumed a rather simple model of a quantum computer.
Every instruction had an execution time of~$1$ and communication latencies between \ac{qpu} and CPU were omitted.
In reality, CPUs execute an instruction much faster than QPUs, and the execution time for one instruction in a QPU varies greatly with the qubit technology used.
Additionally, the execution time of one-qubit and multiple-qubit gates often differs~\cite{ibm_fez_infos,ibm_marrakesh_infos,ibm_torino_infos,ionq_aria}.
An additional consideration is the fact that quantum gates can be executed in parallel if they operate on distinct qubits.
Succeeding work could examine in more detail how this affects quantum-classical optimization.

In \Cref{sec:related-work} we mentioned that there is research on circuit optimization focused solemnly on the quantum circuit.
Combining these quantum circuit optimizations with the optimizations we derived from classical procedures is another non-trivial problem we strive to investigate in the future.

\section{Conclusion}
\label{sec:conclusion}

The aim of this work was to explore optimization strategies for heterogeneous calculations, where a quantum processing unit interacts with a classical device in near-real time. 
While much research has focused on optimizing quantum circuits, little attention has been given to the classical component in real-time quantum-classical calculations. 
Additionally, there has been a lack of established metrics for evaluating the performance of quantum-classical hybrid programs.

In this context, our work makes a significant contribution by introducing three performance metrics—wall time, quantum instruction number, and quantum calculation time—to assess the efficiency of hybrid programs. 
Furthermore, we proposed seven optimization operations for heterogeneous Quil programs, including constant propagation, live-variable analysis, constant folding, dead code elimination, finding hybrid dependencies, instruction reordering, and latest possible quantum execution.

Through practical evaluation, we demonstrated that these operations can indeed optimize programs based on our proposed metrics.
This work paves the way for the development of new hybrid optimization techniques, enhancing the potential for real-time collaboration between quantum and classical devices.
By providing both the foundational metrics and optimization operations, we believe this research opens up exciting avenues for future advancements in hybrid quantum computing.

\section*{Acknowledgment}

Generative AI (GitHub Copilot) was used while working on the code corresponding to this work~\cite{supplementary_github}.
It was used as an integrated tool in the integrated development environments (IDEs).
It has been used to assist during code-writing, helped writing documentation strings, deciding on variable/function names, wrote first drafts of some methods/functions, for formulation suggestions, and for Markdown formatting.

\bibliographystyle{IEEEtran}
\bibliography{references}

% Generated by IEEEtran.bst, version: 1.14 (2015/08/26)
\begin{thebibliography}{10}
\providecommand{\url}[1]{#1}
\csname url@samestyle\endcsname
\providecommand{\newblock}{\relax}
\providecommand{\bibinfo}[2]{#2}
\providecommand{\BIBentrySTDinterwordspacing}{\spaceskip=0pt\relax}
\providecommand{\BIBentryALTinterwordstretchfactor}{4}
\providecommand{\BIBentryALTinterwordspacing}{\spaceskip=\fontdimen2\font plus
\BIBentryALTinterwordstretchfactor\fontdimen3\font minus
  \fontdimen4\font\relax}
\providecommand{\BIBforeignlanguage}[2]{{%
\expandafter\ifx\csname l@#1\endcsname\relax
\typeout{** WARNING: IEEEtran.bst: No hyphenation pattern has been}%
\typeout{** loaded for the language `#1'. Using the pattern for}%
\typeout{** the default language instead.}%
\else
\language=\csname l@#1\endcsname
\fi
#2}}
\providecommand{\BIBdecl}{\relax}
\BIBdecl

\bibitem{Remme2025}
L.~Remme, ``Optimization strategies for quantum computers in distributed
  systems,'' Master's thesis, Heinrich Heine University Düsseldorf, 2 2025.

\bibitem{qiskit_quantum_circuit}
\BIBentryALTinterwordspacing
IBMQuantum. (2024) {circuit} | {IBM Quantum Computing}. [Online]. Available:
  \url{https://docs.quantum.ibm.com/api/qiskit/circuit}
\BIBentrySTDinterwordspacing

\bibitem{pennylane_quantum_circuit}
\BIBentryALTinterwordspacing
Xanadu. (2025) Quantum circuits -- pennylane 0.40.0 documentation. [Online].
  Available:
  \url{https://docs.pennylane.ai/en/stable/introduction/circuits.html}
\BIBentrySTDinterwordspacing

\bibitem{cirq_circuits}
\BIBentryALTinterwordspacing
G.~Q. AI. (2025) Circuits | {Cirq} | {G}oogle {Q}uantum {AI}. [Online].
  Available: \url{https://quantumai.google/cirq/build/circuits}
\BIBentrySTDinterwordspacing

\bibitem{knill1996conventions}
\BIBentryALTinterwordspacing
E.~Knill, ``Conventions for {Q}uantum {P}seudocode,'' Los Alamos National Lab.
  (LANL), Los Alamos, NM (United States), Tech. Rep., 06 1996. [Online].
  Available: \url{https://www.osti.gov/biblio/366453}
\BIBentrySTDinterwordspacing

\bibitem{bennett1993teleporting}
C.~H. Bennett, G.~Brassard, C.~Cr{\'e}peau, R.~Jozsa, A.~Peres, and W.~K.
  Wootters, ``Teleporting an {U}nknown {Q}uantum {S}tate via {D}ual {C}lassical
  and {E}instein-{P}odolsky-{R}osen {C}hannels,'' \emph{Physical Review
  Letters}, vol.~70, pp. 1895--1899, 1993.

\bibitem{lubinski2022advancing}
T.~Lubinski, C.~Granade, A.~Anderson, A.~Geller, M.~Roetteler, A.~Petrenko, and
  B.~Heim, ``Advancing hybrid quantum--classical computation with real-time
  execution,'' \emph{Frontiers in Physics}, vol.~10, 2022.

\bibitem{paetznick2013repeat}
A.~Paetznick and K.~M. Svore, ``Repeat-{U}ntil-{S}uccess: {N}on-deterministic
  decomposition of single-qubit unitaries,'' \emph{Quantum Information \&
  Computation}, vol.~14, no. 15–16, pp. 1277--1301, Nov. 2014.

\bibitem{fu2023proto}
P.~Fu, K.~Kishida, N.~J. Ross, and P.~Selinger, ``{Proto-Quipper} with
  {D}ynamic {L}ifting,'' \emph{Proceedings of the ACM on Programming
  Languages}, vol.~7, no.~11, pp. 309--334, Jan. 2023.

\bibitem{dobvsivcek2007arbitrary}
M.~Dobšíček, G.~Johansson, V.~Shumeiko, and G.~Wendin, ``Arbitrary accuracy
  iterative quantum phase estimation algorithm using a single ancillary qubit:
  {A} two-qubit benchmark,'' \emph{Physical Review A}, vol.~76, 9 2007.

\bibitem{fu2021quingo}
X.~Fu, J.~Yu, X.~Su, H.~Jiang, H.~Wu, F.~Cheng, X.~Deng, J.~Zhang, L.~Jin,
  Y.~Yang \emph{et~al.}, ``Quingo: {A} {P}rogramming {F}ramework for
  {H}eterogeneous {Q}uantum-{C}lassical {C}omputing with {NISQ} {F}eatures,''
  \emph{ACM Transactions on Quantum Computing}, vol.~2, 2021.

\bibitem{rossi2022using}
M.~Rossi, L.~Asproni, D.~Caputo, S.~Rossi, A.~Cusinato, R.~Marini, A.~Agosti,
  and M.~Magagnini, ``Using {S}hor’s algorithm on near term {Q}uantum
  computers: a reduced version,'' \emph{Quantum Machine Intelligence}, vol.~4,
  no.~18, Jul. 2022.

\bibitem{beauregard2002circuit}
S.~Beauregard, ``Circuit for {S}hor's algorithm using 2n+ 3 qubits,''
  \emph{Quantum Information \& Computation}, vol.~3, no.~2, p. 175–185, Mar.
  2003.

\bibitem{knill2004fault}
E.~Knill, ``{F}ault-{T}olerant {P}ostselected {Q}uantum {C}omputation:
  {S}chemes,'' \emph{arXiv preprint quant-ph/0402171}, 2004.

\bibitem{bravyi2005universal}
S.~Bravyi and A.~Kitaev, ``Universal quantum computation with ideal {C}lifford
  gates and noisy ancillas,'' \emph{Physical Review A}, vol.~71, 2005.

\bibitem{fowler2012surface}
A.~G. Fowler, M.~Mariantoni, J.~M. Martinis, and A.~N. Cleland, ``Surface
  codes: {T}owards practical large-scale quantum computation,'' \emph{Physical
  Review A}, vol.~86, 9 2012.

\bibitem{karuppasamy2024quantum}
K.~{Karuppasamy}, V.~{Puram}, S.~{Johnson}, and J.~P. {Thomas}, ``Quantum
  {C}ircuit {O}ptimization: {C}urrent trends and future direction,''
  \emph{arXiv e-prints}, p. arXiv:2408.08941, Aug. 2024.

\bibitem{fosel2021quantum}
T.~{F{\"o}sel}, M.~{Yuezhen Niu}, F.~{Marquardt}, and L.~{Li}, ``Quantum
  circuit optimization with deep reinforcement learning,'' \emph{arXiv
  e-prints}, p. arXiv:2103.07585, Mar. 2021.

\bibitem{ruiz2024quantum}
F.~J. Ruiz, T.~Laakkonen, J.~Bausch, M.~Balog, M.~Barekatain, F.~J. Heras,
  A.~Novikov, N.~Fitzpatrick, B.~Romera-Paredes, J.~van~de Wetering
  \emph{et~al.}, ``Quantum {C}ircuit {O}ptimization with {A}lpha{T}ensor,''
  \emph{arXiv e-prints}, pp. arXiv--2402, 2024.

\bibitem{li2024quarl}
\BIBentryALTinterwordspacing
Z.~Li, J.~Peng, Y.~Mei, S.~Lin, Y.~Wu, O.~Padon, and Z.~Jia, ``Quarl: {A}
  {L}arning-{B}ased {Q}uantum {C}ircuit {O}ptimizer,'' \emph{Proceedings of the
  ACM on Programming Languages}, vol.~8, no. OOPSLA1, Apr. 2024. [Online].
  Available: \url{https://doi.org/10.1145/3649831}
\BIBentrySTDinterwordspacing

\bibitem{smith2016practical}
R.~S. {Smith}, M.~J. {Curtis}, and W.~J. {Zeng}, ``A {P}ractical {Q}uantum
  {I}nstruction {S}et {A}rchitecture,'' \emph{arXiv e-prints}, p.
  arXiv:1608.03355, Aug. 2016.

\bibitem{van2023optimising}
J.~van~de Wetering and M.~Amy, ``Optimising {T}-count is np-hard,'' \emph{arXiv
  e-prints}, Sep. 2023.

\bibitem{herr2017optimization}
D.~Herr, F.~Nori, and S.~J. Devitt, ``Optimization of lattice surgery is
  {NP}-hard,'' \emph{npj Quantum Information}, vol.~3, no.~35, 2017.

\bibitem{van2024optimal}
J.~van~de Wetering, R.~Yeung, T.~Laakkonen, and A.~Kissinger, ``Optimal
  compilation of parametrised quantum circuits,'' \emph{arXiv e-prints}, Jan.
  2024.

\bibitem{smith2020open}
R.~S. Smith, E.~C. Peterson, M.~G. Skilbeck, and E.~J. Davis, ``An open-source,
  industrial-strength optimizing compiler for quantum programs,'' \emph{Quantum
  Science and Technology}, vol.~5, no.~4, 7 2020.

\bibitem{nam2018automated}
Y.~Nam, N.~J. Ross, Y.~Su, A.~M. Childs, and D.~Maslov, ``Automated
  optimization of large quantum circuits with continuous parameters,''
  \emph{npj Quantum Information}, vol.~4, no.~1, p.~12, 2018.

\bibitem{bae2020quantum}
J.-H. Bae, P.~M. Alsing, D.~Ahn, and W.~A. Miller, ``Quantum circuit
  optimization using quantum karnaugh map,'' \emph{Scientific reports},
  vol.~10, no. 15651, 2020.

\bibitem{karnaugh1953map}
M.~Karnaugh, ``The map method for synthesis of combinational logic circuits,''
  \emph{Transactions of the American Institute of Electrical Engineers, Part I:
  Communication and Electronics}, vol.~72, no.~5, pp. 593--599, 1953.

\bibitem{puram2024optimizing}
V.~Puram, K.~Karuppasamy, and J.~P. Thomas, ``Optimizing {Q}uantum {C}ircuits
  {U}sing {A}lgebraic {E}xpressions,'' in \emph{Computational Science -- ICCS
  2024}.\hskip 1em plus 0.5em minus 0.4em\relax Springer Nature Switzerland,
  2024, pp. 268--276.

\bibitem{gheorghiu2022reducing}
V.~Gheorghiu, J.~Huang, S.~M. Li, M.~Mosca, and P.~Mukhopadhyay, ``Reducing the
  {CNOT} {C}ount for {C}lifford+{T} {C}ircuits on {NISQ} {A}rchitectures,''
  \emph{IEEE Transactions on Computer-Aided Design of Integrated Circuits and
  Systems}, vol.~42, no.~6, pp. 1873--1884, 2022.

\bibitem{nagele2024optimizing}
M.~N{\"a}gele and F.~Marquardt, ``Optimizing zx-diagrams with deep
  reinforcement learning,'' \emph{Machine Learning: Science and Technology},
  vol.~5, no.~3, 2024.

\bibitem{coecke2008interacting}
B.~Coecke and R.~Duncan, ``Interacting {Q}uantum {O}bservables,'' in
  \emph{Proceedings of the 35th international colloquium on Automata, Languages
  and Programming, Part II}, 2008, pp. 298--310.

\bibitem{quietschlich2023compiler}
N.~Quetschlich, L.~Burgholzer, and R.~Wille, ``Compiler {O}ptimization for
  {Q}uantum {C}omputing {U}sing {R}einforcement {L}earning,'' in \emph{2023
  60th ACM/IEEE Design Automation Conference (DAC)}, 2023, pp. 1--6.

\bibitem{chen2022qcir}
M.~Chen, Y.~Zhang, Y.~Li, Z.~Wang, J.~Li, and X.~Li, ``Qcir: {P}attern
  {M}atching {B}ased {U}niversal {Q}uantum {C}ircuit {R}ewriting {F}ramework,''
  in \emph{ICCAD '22: Proceedings of the 41st IEEE/ACM International Conference
  on Computer-Aided Design}, 2022, pp. 1--8.

\bibitem{iten2022exact}
R.~Iten, R.~Moyard, T.~Metger, D.~Sutter, and S.~Woerner, ``Exact and
  {P}ractical {P}attern {M}atching for {Q}uantum {C}ircuit {O}ptimization,''
  \emph{ACM Transactions on Quantum Computing}, vol.~3, no.~1, pp. 1--41, 2022.

\bibitem{gemeinhardt2025gequpi}
F.~Gemeinhardt, S.~Klikovits, and M.~Wimmer, ``Gequpi: {Q}uantum {P}rogram
  {I}mprovement with {M}ulti-{O}bjective {G}enetic {P}rogramming,''
  \emph{Journal of Systems and Software}, vol. 219, 2025.

\bibitem{wei2021genetic}
L.~Wei, Z.~Ma, Y.~Cheng, and Q.~Liu, ``Genetic {A}lgorithm {B}ased {Q}uantum
  {C}ircuits {O}ptimization for {Q}uantum {C}omputing {S}imulation,'' in
  \emph{2021 12th International Conference on Information, Intelligence,
  Systems \& Applications (IISA)}.\hskip 1em plus 0.5em minus 0.4em\relax IEEE,
  2021, pp. 1--8.

\bibitem{pointing2024optimizing}
J.~Pointing, O.~Padon, Z.~Jia, H.~Ma, A.~Hirth, J.~Palsberg, and A.~Aiken,
  ``Quanto: Optimizing quantum circuits with automatic generation of circuit
  identities,'' \emph{Quantum Science and Technology}, vol.~9, no.~4, 2024.

\bibitem{xu2022quartz}
M.~Xu, Z.~Li, O.~Padon, S.~Lin, J.~Pointing, A.~Hirth, H.~Ma, J.~Palsberg,
  A.~Aiken, U.~A. Acar, and J.~Zhihao, ``Quartz: superoptimization of {Q}uantum
  circuits,'' in \emph{Proceedings of the 43rd ACM SIGPLAN International
  Conference on Programming Language Design and Implementation}, 2022, pp.
  625--640.

\bibitem{paykin2023pcoast}
J.~Paykin, A.~T. Schmitz, M.~Ibrahim, X.-C. Wu, and A.~Y. Matsuura, ``Pcoast:
  {A} {P}auli-{B}ased {Q}uantum {C}ircuit {O}ptimization {F}ramework,'' in
  \emph{2023 IEEE International Conference on Quantum Computing and Engineering
  (QCE)}, vol.~1.\hskip 1em plus 0.5em minus 0.4em\relax IEEE, 2023, pp.
  715--726.

\bibitem{hietala2021verified}
K.~Hietala, R.~Rand, S.-H. Hung, X.~Wu, and M.~Hicks, ``A {V}erified
  {O}ptimizer for {Q}uantum {C}ircuits,'' \emph{Proceedings of the ACM on
  Programming Languages}, vol.~5, Jan. 2021.

\bibitem{coq2024}
\BIBentryALTinterwordspacing
{The Coq Development Team}. (2024, Sep.) The coq proof assistant. [Online].
  Available: \url{https://doi.org/10.5281/zenodo.11551307}
\BIBentrySTDinterwordspacing

\bibitem{wille2024qdmi}
R.~Wille, L.~Schmid, Y.~Stade, J.~Echavarria, M.~Schulz, L.~Schulz, and
  L.~Burgholzer, ``Qdmi - quantum device management interface:
  Hardware-software interface for the munich quantum software stack,'' in
  \emph{QCE}, vol.~02, 2024, pp. 573--574.

\bibitem{almudever2024designing}
C.~G. Almudever, R.~Wille, F.~Sebastiano, N.~Haider, and E.~Alarcon, ``From
  designing quantum processors to large-scale quantum computing systems,'' in
  \emph{DATE}, 2024, pp. 1--10.

\bibitem{kaya2024software}
E.~Kaya, J.~Echavarria, M.~N. Farooqi, A.~Swierkowska, P.~Hopf, B.~Mete,
  L.~Burgholzer, R.~Wille, L.~Schulz, and M.~Schulz, ``A software platform to
  support disaggregated quantum accelerators,'' in \emph{SC24-W: Workshops of
  the International Conference for High Performance Computing, Networking,
  Storage and Analysis}, 2024, pp. 1646--1653.

\bibitem{basermann2025quantum}
\BIBentryALTinterwordspacing
A.~Basermann, M.~Epping, B.~Fauseweh, M.~Felderer, E.~Lobe,
  M.~R{\"{o}}hrig{-}Z{\"{o}}llner, G.~Schmiedinghoff, P.~K. Schuhmacher,
  Y.~Setyawati, and A.~Weinert, ``Quantum software ecosystem design,'' in
  \emph{Software Engineering 2025 Companion Proceedings, Fachtagung des
  GI-Fachbereichs Softwaretechnik, Karlsruhe, Germany, February 24-28, 2025},
  K.~Feichtinger, L.~Sonnleithner, and H.~Hajiabadi, Eds.\hskip 1em plus 0.5em
  minus 0.4em\relax Gesellschaft f{\"{u}}r Informatik e.V., 2025, p.~22.
  [Online]. Available: \url{https://doi.org/10.18420/se2025-ws-22}
\BIBentrySTDinterwordspacing

\bibitem{ThunigJohannfunkeWangEtAl2024}
R.~Thunig, M.~Johannfunke, T.~Wang, and H.~Schirmeier, ``One flag to rule them
  all? on the quest for compiler optimizations to improve fault tolerance,'' in
  \emph{2024 19th European Dependable Computing Conference (EDCC)}.\hskip 1em
  plus 0.5em minus 0.4em\relax IEEE, 2024, pp. 33--40.

\bibitem{LiuRenXieEtAl2024}
J.~Liu, J.~Ren, J.~Xie, J.~Fang, and T.~Wang, ``Enhancing compiler optimization
  with reinforcement learning and monte carlo tree search,'' in
  \emph{Proceedings of the 36th International Conference on Software
  Engineering and Knowledge Engineering}, ser. SEKE2024, vol. 2024.\hskip 1em
  plus 0.5em minus 0.4em\relax KSI Research Inc., Oct. 2024, pp. 146--151.

\bibitem{AhmedFahimUlHaqueRazaKhanEtAl2024}
H.~Ahmed, M.~Fahim Ul~Haque, H.~Raza~Khan, G.~Nadeem, K.~Arshad, K.~Assaleh,
  and P.~Cesar~Santos, ``Selecting the best compiler optimization by adopting
  natural language processing,'' \emph{IEEE Access}, vol.~12, pp.
  121\,700--121\,711, 2024.

\bibitem{ashouri2018survey}
A.~H. Ashouri, W.~Killian, J.~Cavazos, G.~Palermo, and C.~Silvano, ``A {S}urvey
  on {C}ompiler {A}utotuning using {M}achine {L}earning,'' \emph{ACM Computing
  Surveys (CSUR)}, vol.~51, no.~96, pp. 1--42, 9 2018.

\bibitem{triantafyllis2003compiler}
S.~Triantafyllis, M.~Vachharajani, N.~Vachharajani, and D.~I. August,
  ``Compiler optimization-space exploration,'' in \emph{International Symposium
  on Code Generation and Optimization, 2003. CGO 2003.}\hskip 1em plus 0.5em
  minus 0.4em\relax IEEE, 3 2003, pp. 204--215.

\bibitem{aho2007compilers}
A.~V. Aho, M.~S. Lam, R.~Sethi, and J.~D. Ullman, \emph{Compilers:
  {P}rinciples, {T}echniques and {T}ools}, 2nd~ed.\hskip 1em plus 0.5em minus
  0.4em\relax Pearson Education, 2007.

\bibitem{computing2019pyquil}
\BIBentryALTinterwordspacing
{Rigetti Computing}. (2019, Jan.) {pyQuil Documentation Release 2.3.0}.
  [Online]. Available:
  \url{https://readthedocs.org/projects/pyquil/downloads/pdf/v2.3.0/}
\BIBentrySTDinterwordspacing

\bibitem{qvm_quil_project}
\BIBentryALTinterwordspacing
quil lang. (2024) qvm: {A} {H}igh-{P}erformance {Q}uantum {V}irtual {M}achine.
  [Online]. Available: \url{https://github.com/quil-lang/qvm}
\BIBentrySTDinterwordspacing

\bibitem{assemblyManual}
\BIBentryALTinterwordspacing
{Sun Microsystems}. (1995) {x86 Assembly Language Reference Manual}. [Online].
  Available: \url{https://docs.oracle.com/cd/E19641-01/802-1948/802-1948.pdf}
\BIBentrySTDinterwordspacing

\bibitem{quil_language_specification}
\BIBentryALTinterwordspacing
{Robert S. Smith; Rigetti \& Co. Inc.; and contributors}. (2021) {Quil
  Specification}. Version 2021.1. [Online]. Available:
  \url{https://quil-lang.github.io/}
\BIBentrySTDinterwordspacing

\bibitem{supplementary_github}
\BIBentryALTinterwordspacing
lirem101. Github - {LiRem101/parser-analyser}: {A}nalysation and optimization
  of {Q}uil programs. [Online]. Available:
  \url{https://github.com/LiRem101/parser-analyser}
\BIBentrySTDinterwordspacing

\bibitem{allen1970control}
\BIBentryALTinterwordspacing
F.~E. Allen, ``Control flow analysis,'' \emph{ACM SIGPLAN Notices}, vol.~5, pp.
  1--–19, Jul. 1970. [Online]. Available:
  \url{https://doi.org/10.1145/390013.808479}
\BIBentrySTDinterwordspacing

\bibitem{cooper2022engineering}
K.~D. Cooper and L.~Torczon, \emph{Engineering a {C}ompiler}.\hskip 1em plus
  0.5em minus 0.4em\relax Morgan Kaufmann, 2006.

\bibitem{peres2004quantum}
A.~Peres and D.~R. Terno, ``Quantum {I}nformation and {R}elativity {T}heory,''
  \emph{Reviews of Modern Physics}, vol.~76, pp. 93--123, 1 2004.

\bibitem{gottesman1998theory}
D.~Gottesman, ``Theory of fault-tolerant quantum computation,'' \emph{Physical
  Review A}, vol.~57, pp. 127--137, 1 1998.

\bibitem{wootters1982single}
W.~K. Wootters and W.~H. Zurek, ``A single quantum cannot be cloned,''
  \emph{Nature}, vol. 299, pp. 802--803, 1982.

\bibitem{dieks1982communication}
\BIBentryALTinterwordspacing
D.~Dieks, ``Communication by {EPR} devices,'' \emph{Physics Letters A},
  vol.~92, no.~6, pp. 271--272, 1982. [Online]. Available:
  \url{https://www.sciencedirect.com/science/article/pii/0375960182900846}
\BIBentrySTDinterwordspacing

\bibitem{ibm_fez_infos}
\BIBentryALTinterwordspacing
IBMQuantum. (2024) ibm\_fez. [Online]. Available:
  \url{https://quantum.ibm.com/services/resources?system=ibm\_fez}
\BIBentrySTDinterwordspacing

\bibitem{ibm_marrakesh_infos}
\BIBentryALTinterwordspacing
------. (2024) ibm\_marrakesh. [Online]. Available:
  \url{https://quantum.ibm.com/services/resources?system=ibm\_marrakesh}
\BIBentrySTDinterwordspacing

\bibitem{ibm_torino_infos}
\BIBentryALTinterwordspacing
------. (2024) ibm\_torino. [Online]. Available:
  \url{https://quantum.ibm.com/services/resources?system=ibm\_torino}
\BIBentrySTDinterwordspacing

\bibitem{ionq_aria}
\BIBentryALTinterwordspacing
I.~IonQ. (2024) Ionq {A}ria: {P}ractical {P}erformance. [Online]. Available:
  \url{https://ionq.com/resources/ionq-aria-practical-performance}
\BIBentrySTDinterwordspacing

\end{thebibliography}

\end{document}